# Observation of the crossover from quantum fluxoid to half-quantum fluxoid in a chiral superconducting device

Masashi Tokuda[1], Fumiya Matsumoto[1], Noriaki Maeda[1], Tomo Higashihara[1], Mai Nakao[1], Mori Watanabe[1], Sanghyun Lee[1], Ryoya Nakamura[1], Masaki Maeda[1], Nan Jiang[1,2,3], Di Yue[4,5], Hideki Narita[6], Kazushi Aoyama[7], Takeshi Mizushima[8], Jun-ichiro Ohe[9], Teruo Ono[6,10], Xiaofeng Jin[4,5], Kensuke Kobayashi[11], and Yasuhiro Niimi[1,2,3*]

[1]Department of Physics, Graduate School of Science, Osaka University, Toyonaka, Osaka 560-0043, Japan.

[2]Center for Spintronics Research Network, Graduate School of Engineering Science, Osaka University, Toyonaka, Osaka 560-8531, Japan.

[3]Spintronics Research Network Division, Institute for Open and Transdisciplinary Research Initiatives, Osaka University, Suita, Osaka, 565-0871, Japan.

[4]State Key Laboratory of Surface Physics and Department of Physics, Fudan University, Shanghai 200433, People's Republic of China.

[5]Collaborative Innovation Center of Advanced Microstructures, Fudan University, Shanghai 200433, People's Republic of China.

[6]Institute for Chemical Research, Kyoto University, Uji, Kyoto 611-0011, Japan

[7]Department of Earth and Space Science, Graduate School of Science, Osaka University, Toyonaka, Osaka 560-0043, Japan.

[8]Department of Materials Engineering Science, Osaka University, Toyonaka, Osaka 560-8531, Japan.

[9]Department of Physics, Toho University, Funabashi, Chiba 274-8510, Japan

[10]Center for Spintronics Research Network, Institute for Chemical Research, Kyoto University, Uji, Kyoto 611-0011, Japan

[11]Institute for Physics of Intelligence and Department of Physics, Graduate School of Science, The University of Tokyo, Bunkyo-ku, Tokyo 113-0033, Japan.

*e-mail: niimi@phys.sci.osaka-u.ac.jp

**Topological superconductors are one of the intriguing material groups from the viewpoint of not only condensed matter physics but also industrial application such as quantum computers based on Majorana fermion. For the real application, developments of the thin-film topological superconductors are highly desirable. Bi/Ni bilayer is a possible candidate for thin-film chiral superconductors where the time-reversal symmetry is broken. Here we report the phase shift of resistance oscillations by half flux quantum in a ring-shaped**

**device of epitaxial Bi/Ni bilayer induced by a small magnetic field through the ring. The half quantum fluxoid can be a decisive evidence for unconventional superconductors where the superconducting order parameter has an internal degree of freedom. The present result provides a functional operating principle for quantum devices where the phase of the supercurrent can be shifted by $\pi$ with a small magnetic field, based on the internal degree of freedom possessed by topological superconductivity.**

**Introduction**

Chiral superconductors, which are a type of topological superconductors, show notable physical properties (*1–15*): for example, the demonstration of spontaneous time-reversal symmetry-breaking and the existence of Majorana fermion. To take advantage of these fascinating properties for application, the chiral superconductors have to be fabricated into sub-micrometer devices. However, most of the candidates of chiral superconductors are in bulk form, which is not suitable for nanofabrication and also for mass production as in the case of semiconductor devices. Bi/Ni bilayer, illustrated in Fig. 1(A), is a candidate of thin-film chiral superconductor where the time-reversal symmetry-breaking has been detected by Kerr rotation spectroscopy (*13*) only below the superconducting transition temperature $T_c = 4.2$ K.

In order to evaluate the superconducting order parameter of Bi/Ni, we have measured resistance oscillations (*16*, *17*) in a ring-shaped device. When a magnetic field is applied through a superconducting ring, the phase difference of wave function around the ring must be an integer multiple of $2\pi$. As a result, the magnetic flux $\phi$ inside the ring is quantized by the unit of magnetic flux quantum $\phi_0$, i.e., $\phi = n\phi_0$, where $n$ is the integer and $\phi_0 = (h/2e)$ ($h$ is the Planck constant and $e$ is the elementary charge). In this situation, $T_c$ oscillates as a function of the external magnetic field with a period of $\phi_0$. This is called Little-Parks (LP) oscillations, where the free energy takes minima at $\phi = n\phi_0$. The oscillations of $T_c$ are also reflected in the magnetic field dependence of resistance, known as quantum fluxoids (QFs)-type LP oscillations. In general, the QFs-type LP oscillations are realized in conventional spin-singlet $s$-wave superconductors. Meanwhile, in unconventional superconductors such as a spin-triplet chiral $p$-wave superconductor, half-quantum fluxoids (HQFs)-type oscillations are observable (*18*, *19*), where the phase of oscillations is shifted by $\phi_0/2$ from the QFs-type oscillations: in short, $\phi$ inside the ring is quantized by $(n + 1/2)\phi_0$ (*3*, *20*–*25*). Therefore, the observation of HQFs can be an evidence of unconventional superconductivity.

Here we demonstrate the crossover from QFs-type to HQFs-type resistance oscillations in ring-shaped devices fabricated with an epitaxially-grown Bi(35 nm)/Ni(2 nm) superconducting thin film by applying a small magnetic field (~8 mT) through the rings. The result indicates that the *in-situ* phase modulation can be realized in the Bi/Ni ring devices by the small magnetic field and is due to the internal degree of freedom of the order parameter, unlike the cases of corner junction in $d$-wave superconductors (*26*), Josephson $\pi$-junctions in polycrystalline devices (*18*), and half-metallic ferromagnet/superconductor heterojunction (*27*). The ability to change the phase by $\pi$ as needed could be incorporated into superconducting quantum circuits to enrich the computational capabilities.

**Results**

The ring device fabricated with an epitaxial Bi/Ni bilayer is illustrated in Fig. 1(B). The inset of Fig. 1(C) shows a scanning electron microscope (SEM) image of the device. The temperature dependence of the resistance is shown in Fig. 1(C). A sharp superconducting transition is clearly observed at about 4 K as in the case of epitaxial Bi/Ni bilayer films (*15*). This indicates that very little damage is given by the device fabrication processes (see Materials and Methods for the detailed device fabrication processes and Supplementary Materials).

As mentioned in the introduction, LP oscillations are discussed as a modulation of $T_c$: $T_c$ oscillates as a function of magnetic flux through a ring and the critical current $I_c$ becomes almost zero since the temperature is close to $T_c$. In the present case, however, it is difficult to precisely measure the variation of $T_c$ just below 4 K. Therefore, we applied a relatively large current to suppress the superconductivity at our lowest temperature ($T \approx 2.4$ K) and measured resistance oscillations. This situation is different from the original LP oscillation, because the pair-breaking due to the current $I$ is the first order transition while the one due to $T$ is the second order transition. Based on a recent theoretical work (28), the LP oscillation can be generalized to lower temperatures and finite critical currents. In such a case, the LP-like quantum oscillation in $I_c$ can also be expected at $T \gtrsim 0.6T_c$ as a function of magnetic flux. In Supplementary Materials, we have explained the LP-like oscillation in $I_c$ in more detail. We first measured dc $I$–$V$ characteristics of the epitaxially-grown Bi/Ni ring-shaped device at different magnetic fields and confirmed that the critical current $I_c$ oscillates as a function of magnetic field, as expected in the theory (28) (see figs. S7 and S8 in Supplementary Materials). We next fixed the measurement temperature at $T = 2.47(\pm 0.01)$ K and also the measurement current ($I_m$) in the vicinity of $I_c$ where the temperature is stable and the superconductivity is weak enough to observe resistance oscillations. In such a situation, the LP-like oscillations can be detected as voltage $V$ oscillations, more practically resistance ($R = V/I_m$) changes. Hereafter the LP-like oscillations are simply referred to as resistance oscillations. In order to evaluate the phase of resistance oscillations, a precise magnetic field measurement with a resolution of $\approx 0.01$ mT is also required. For this purpose, we used an electromagnet with a Hall sensor to obtain the precise value of magnetic field and measured a Nb (typical spin singlet $s$-wave superconductor) ring device (29) simultaneously which was set on the same sample holder as shown in Fig. 1(D). As detailed later on, the Nb device is useful for the zero field correction and also for the evaluation of the superconducting order parameter of epitaxial Bi/Ni.

To measure the resistance oscillations precisely, we have adopted the ac lock-in technique with the external magnetic field $H$ perpendicular to the basal plane. The typical result is shown in Fig. 2. Figures 2(A) and 2(B) show the magnetic field dependence of resistances of the epitaxially-grown Bi/Ni and Nb ring devices with almost the same size, simultaneously obtained at $T = 2.47$ K ($\approx 0.6T_c$ for Bi/Ni). Clear oscillations of resistances are observed for both devices. The periods of the oscillations are 0.96 and 0.84 mT for the Bi/Ni and Nb rings, respectively. Those periods are consistent with the value expected from the relation of $\mu_0 H_0 = \phi_0/S$ where $\mu_0$ is the vacuum permeability and $S$ is the effective surface of the ring. As discussed later, the resistance oscillations observed for the epitaxially-grown Bi/Ni ring device are suppressed at about $\pm 8$ mT ($\equiv \mu_0 H^*$) and then recovered above $\mu_0 H^*$. On the other hand, such behavior has not been observed in the reference Nb ring: the amplitude of resistance oscillations is almost constant.

Before evaluating the oscillations in the epitaxially-grown Bi/Ni ring device, the offset of zero magnetic field must be determined. The real zero magnetic field for superconducting rings can be shifted for some reasons such as remnant magnetism of iron core of electromagnet, terrestrial magnetism, and so on. In Figs. 2(A) and 2(B), it is slightly shifted from $\mu_0 H = 0$ obtained with the Hall sensor. Even though it is shifted, resistance oscillations with two different $\mu_0 H_0$ should take minima or maxima at the same time at the real zero field because of the relations of $H = nH_0$ for QF and $H = (n + 1/2)H_0$ for HQF. In this measurement, the real zero field was at about $-0.2$ mT as indicated by the vertical dotted line. Thus, both the resistance oscillations show QFs-type oscillations, i.e., $H = nH_0$ in the low magnetic field region.

We focus on the resistance oscillations obtained for the epitaxial Bi/Ni ring. In Fig. 2(C), we show the magnetic field dependence of resistance of epitaxial Bi/Ni ring. We note that the data points are the same as in Fig. 2(A), but the magnetic field is shifted by +0.2 mT based on the above discussion and the horizontal axis is normalized by $\mu_0 H_0$. As mentioned above, the epitaxial Bi/Ni ring shows QFs-type oscillations in the vicinity of the zero magnetic field, while the phase of oscillations is shifted by $\mu_0 H_0/2$ above $\mu_0 H^*$. In other words, the crossover from QFs to HQFs takes place at $\pm\mu_0 H^*$. We performed the same experiment at $T = 3.6$ K ($\approx 0.9 T_c$) (see Supplementary Materials and Methods). In this case, only the QFs-type oscillations were observed as shown in Fig. 2(D).

We have checked the reproducibility of the crossover for the same device (device #1) in different cooling runs and with different $I_m$ values (see Fig. 2(E)), and also for other ring devices with a similar diameter (device #2) in Fig. 2(F) and with a larger diameter (device #3) which corresponds to an oscillation period of $\mu_0 H_0 \approx 0.6$ mT shown in fig. S13 in Supplementary Materials. For all the devices, the crossover from QFs-type to HQFs-type resistance oscillations takes place at $T \approx 0.6 T_c$ at about $\pm 8$ mT, showing that $\mu_0 H^*$ is independent of the device size. On the other hand, such a crossover does not appear at $T \approx 0.9 T_c$ as shown in fig. S15 (device #5) as well as in Fig. 2(D) (device #1).

We have also performed control experiments. In Fig. 3(A), we show resistance oscillations of a polycrystalline Bi/Ni ring, where the domain size should be of the order of tens nm (*18*), measured at $T \approx 0.6T_c$. However, unlike the case of polycrystalline $\beta$-Bi$_2$Pd rings where HQFs-type LP oscillations were observed near $T_c$ (*18*), only the QFs-type oscillations were detected. We have checked four different polycrystalline Bi/Ni ring devices, but none of them has shown HQFs (see fig. S16 in Supplementary Materials). This result suggests that the domain size of Bi/Ni (*13*) would be key for the observation of the crossover from QFs- to HQFs-type resistance oscillations. Next, we prepared Nb(15 nm)/Ni(3 nm) films on an MgO(001) substrate and patterned ring-shaped devices. A typical result is shown in Fig. 3(B). No crossover from QFs- to HQFs-type oscillations was observed at $T \approx 0.6T_c$ (see also fig. S17 in Supplementary Materials for the reproducibility). All the results indicate that the crossover from QFs- to HQFs-type resistance oscillations is typical of epitaxial Bi/Ni rings.

**Discussion**

We now discuss the mechanism of the crossover from QFs- to HQFs-type resistance oscillations. One possibility is the enhancement of magnetic flux through the ring induced by the stray field of perpendicularly polarized Ni. However, this possibility can be excluded because the easy axis of the magnetic moment of Ni is in-plane and the external magnetic field to polarize the magnetic moment along the out-of-plane direction is about 400 mT (*30*) which is two orders of magnitude larger than $\mu_0 H^*$. Thus, the magnetization of the Ni layer at $\mu_0 H^*$ is still aligned along the in-plane direction and does not pass through the ring. This has been confirmed in our micromagnetic simulations (see fig. S18 in Supplementary Materials). There is also the possibility that magnetic vortices induced by the external magnetic field or the injected current are pinned (or penetrated) inside the ring. However, this can also be ruled out because the crossover from QFs- to HQFs-type resistance oscillations always occurs at the same magnetic field, $\mu_0 H^* \sim \pm 8$ mT, even with different $I_\mathrm{m}$ values (Fig. 2(E)), different ring devices (Fig. 2(F) and fig. S13 in Supplementary Materials), and bi-directional field sweeping (fig. S12 in Supplementary Materials).

HQFs-type resistance oscillations have been observed in Sr$_2$RuO$_4$ (*19*, *31*), $\alpha$-BiPd (*25*), and $\beta$-Bi$_2$Pd (*18*, *32*) devices. In the case of Sr$_2$RuO$_4$, the external in-plane magnetic field is essential to realize the HQFs-state (*19*). In the present case, on the other hand, we apply only the perpendicular magnetic field through the ring. According to Ref. 31, the chiral domain walls can be weak links of Josephson junctions which might explain the resistance oscillations in the epitaxial Bi/Ni rings. However, such chiral domain walls should not be stable because the width of the ring (500 nm) is much longer than the coherence length ($\xi_\parallel \approx 22$ nm, see fig. S22 and related discussions in Supplementary Materials) (*31*). Thus, weak links of Josephson junctions would not be formed in the epitaxial Bi/Ni rings. In the latter two cases, the odd number of Josephson $\pi$ junctions is required to explain the HQFs-type LP oscillations. In fact, not all but some of polycrystalline $\alpha$-BiPd and $\beta$-Bi$_2$Pd devices with incidental formation of superconducting weak links could induce the odd number of Josephson $\pi$ junctions and thus showed the HQFs-type LP oscillations even at zero field. In the epitaxial Bi/Ni rings, on the other hand, HQFs-type resistance oscillations have been observed only above $\mu_0 H^*$ for all the investigated devices. In epitaxial $\beta$-Bi$_2$Pd devices (*33*), higher-order QFs-type resistance oscillations have been reported, although there is no crossover from QFs- to HQFs-type resistance oscillations. Therefore, the mechanism is different from those for the Sr$_2$RuO$_4$, $\alpha$-BiPd, and $\beta$-Bi$_2$Pd devices.

Another possibility is due to the superconducting order parameter. There have been several reports on the gap structures in the superconducting Bi/Ni bilayer: for example, chiral $d$-wave superconductor (*13, 34*), and effective $p$-wave like superconductors (*35–38*). Some of them suggested that the surface state of Bi/Ni bilayer is essential for unconventional superconductivity (*13, 34, 36*). As demonstrated in our upper critical field measurements (see fig. S20 in Supplementary Materials), however, the bulk component is more dominant for the superconductivity in Bi/Ni, which is consistent with Ref. 35. This result indicates that the surface-originated mechanism such as the Rashba effect would be irrelevant for the crossover from QFs- to HQFs-type resistance oscillations.

The most feasible mechanism of the crossover from QFs- to HQFs-type resistance oscillations is the spin texture in spin-triplet superconducting rings with a bias current. As detailed in Sec. 9 in Supplementary Materials as well as in Refs. 28 and 39, the spin (more precisely **d**-vector) degrees of freedom is essential for resistance oscillations. Just below $T_c$, $I_c$ is finite but almost zero. In such a case, the homogenous **d**-vector texture is energetically stable, leading to QFs-type oscillations even in the triplet pairing, and the superconductivity vanishes under a small electric current and magnetic field. When $T$ is lowered, a much larger current passes through the ring, where the current distributions and **d**-vector textures are not equivalent in the upper and lower arms (*28*). However, QFs-type resistance oscillations are still favorable in all the magnetic flux region (*28*), which is inconsistent with the present experimental results for epitaxial Bi/Ni rings. Here we implement a dipole-type spin-orbit interaction for the A-phase of $^3$He superfluid, which favors a specific type of spin texture in the theoretical model as detailed in Ref. 39 and in Sec. 9 in Supplementary Materials. Furthermore, as demonstrated in the experiments, we consider an effective in-plane magnetic field due to the stray field from the Ni layer (*12*) as well as an out-of-plane external magnetic field $H$. Such a small but finite in-plane stray field from the Ni layer has been confirmed by miscromagnetic simulations (see fig. S18 in Supplementary Materials). In the low field region, the direction of the **d**-vector is spatially uniform around the ring (see the left inset of Fig. 4(A)), resulting in QFs-type resistance oscillations as indicated by the black dotted line on the green solid line in Fig. 4(A). With increasing $H$, a finite difference in the superconducting gap function between spin-up and spin-down triplet components is induced. In such a situation, the direction of the **d**-vector becomes spatially non-uniform around the ring (see the right inset of Fig. 4(A)), resulting in HQFs-type resistance oscillations at higher magnetic fields as indicated by the black dotted line on the red solid line in Fig. 4(A). In addition, at $T \approx 0.9T_c$, only the QFs-type resistance oscillations are expected (see Fig. 4(B)), as demonstrated in the experiments in Fig. 2(D) and fig. S15. Therefore, the experimentally observed crossover from QFs- to HQFs-type resistance oscillations in epitaxial Bi/Ni rings would originate from the spin texture in spin-triplet superconducting rings with a finite spin-orbit coupling under a bias current. The fact that no crossover from QFs- to HQFs-type resistance oscillations is observed either in polycrystalline Bi/Ni or in Nb/Ni is also supportive for our scenario where a single chiral domain with strong spin-orbit coupling in epitaxial Bi/Ni thin films is essential (see Secs. 8 and 9 in Supplementary Materials).

In conclusion, we have demonstrated the phase switching in resistance oscillations using mesoscopic ring-shaped epitaxial Bi/Ni bilayer devices. By applying an external magnetic field through the rings, the crossover from QFs-type to HQFs-type is observed as the resistance oscillations. In order to interpret the crossover, we considered the spin texture based on the A-phase of superfluid $^3$He. We found that the crossover from QFs- to HQFs-type resistance oscillations can be explained by the spin texture in spin-triplet superconducting rings with a finite spin-orbit interaction and bias current. The present results suggest that the epitaxial Bi/Ni bilayer not only possesses a spin-triplet component in its superconductivity but also provides an excellent platform for researches on unconventional superconductivity in nanoscale devices. More importantly, the principle of the phase modulation in resistance oscillations via a magnetic field as small as 8 mT could be incorporated into the present superconducting circuits and enrich the computational capabilities.

**Materials and Methods**

An epitaxial Bi/Ni bilayer was prepared on an MgO(001) substrate by molecule beam epitaxy method (MBE) (*15*). A 2 nm thick Ni film was first deposited on the substrate and then a 35 nm thick Bi film was subsequently deposited without breaking vacuum. The thickness of each layer is optimized so that the bilayer has the highest $T_c$ (*13*, *15*, *40*). The MBE method enables us to obtain the best quality Bi/Ni bilayer: it is well-crystallized at each layer (Bi(110) and Ni(001)) and has a sharp interface between the Bi and Ni layers. This has been confirmed by reflection high-energy electron diffraction patterns and cross-sectional scanning transmission electric microscopy as shown in Ref. 15.

As control samples, we also prepared a polycrystalline Bi/Ni bilayer on a SiO$_2$ substrate and a Nb/Ni bilayer on an MgO(001) substrate. The polycrystalline Bi(35 nm)/Ni(2 nm) film was grown with the same method as the epitaxial one but only the substrate is different. On the other hand, the Nb(15 nm)/Ni(3 nm) film was grown on an MgO substrate by sputtering and subsequent annealing processes. The thicknesses of Nb and Ni were set such that $T_c$ becomes almost the same as the Bi/Ni bilayer.

Using electron beam lithography on a polymethyl-methacrylate including 4% anisole (PMMA A4) resist and Ar milling process, ring-shaped devices are patterned. It should be noted that we have never performed an ordinal heating treatment such as baking process for the PMMA resist throughout the lithography process, in order to avoid mixing Bi and Ni atoms at the layer interface, resulting in a possible formation of $Bi_3Ni$ alloys (*41*). For this purpose, after spin-coating the PMMA resist onto the Bi/Ni bilayer film, the bilayer was put under vacuum (less than ~1 Pa by using a rotary pump) for more than 2 hours to evaporate the solvent of the resist.

To identify the real zero field, we prepared Nb ring devices on a $SiO_2$/Si substrate by using electron beam lithography on a PMMA A4 resist with a standard baking process and dc magnetron sputtering. Although $T_c$ of bulk Nb is about 9 K, $T_c$ of Nb thin film depends on the thickness ($t$). The present Nb ring has $t \approx 30$ nm and $T_c \approx 5$ K (*29*).

For measurements of the resistance oscillations in the ring-shaped devices, standard four-terminal measurements were performed using the ac lock-in technique except for the case of dc $I$-$V$ curve measurements. The experimental setup is also depicted in Fig. 1(B). The devices were cooled down to 2 K by using a $^4$He flow refrigerator. The external magnetic field was applied to the devices perpendicularly by a rotational electromagnet and measured with a Hall sensor. It should be noted that the size of the Hall probe is comparable to that of the sample holder and the uniformity of the magnetic field at the magnet center ($\approx 0.01$ mT/cm) where the two devices are located is well assured. Because it was difficult to keep the temperature near $T_c \approx 4$ K with the $^4$He flow cryostat, measurements of the resistance oscillations at 3.6 K were performed by using a variable temperature insert with a superconducting magnet from Oxford Instruments. Although the stability of the temperature was not as good as with the $^4$He flow cryostat at the lowest temperature, we managed to obtain resistance oscillations.

**Acknowledgments**

We thank K. Iwashita, R. Asama, H. Taniguchi and M. Yokoi for their technical supports. This work was supported by JSPS KAKENHI (Grants Nos. JP16H05964, JP17K18756, JP19H05826, JP20H02557, JP20K03860, JP20H01857, JP20J20229, JP21H01039, JP21K03469, JP22H01221, JP22H04480, and JP22H04481, JP24H00007), JST FOREST (Grant No. JPMJFR2134), the National Basic Research Program of China (Grants Nos. 2015CB921402, 2011CB921802, and 2016YFA0300703), the National Natural Science Foundation of China (Grants Nos. 11374057, 11434003, and 11421404), the Mazda Foundation, Shimadzu Science Foundation, Yazaki Memorial Foundation for Science and Technology, SCAT Foundation, the Asahi glass foundation, and the cooperative Research Project of RIEC, Tohoku University.

**Data and materials availability**

All data needed to evaluate the conclusions in the paper are present in the paper and/or the Supplementary Materials.

**Author Contributions**

M. T. and Y. N. conceived and designed the experiments. M. T., F. M., N. M., T. H., M. N., M. W., S.-H. L., R. N., M. M., N. J., K. K., and Y. N. fabricated devices and performed the transport measurements and analysis. D. Y. and X.-F. J. grew epitaxial and polycrystalline Bi/Ni thin films and H. N. and T. O. grew Nb/Ni thin films. K. A. and T. M. developed the theoretical models. J. O. performed micromagnetic simulations. M. T. and Y. N. wrote the manuscript with the support by K. A. and T. M. All authors discussed the results and commented on the manuscript.

**Corresponding authors**

Correspondence to Yasuhiro Niimi.

**Competing interests**

The authors declare no competing interests.

**Figure captions**

**Fig. 1. Ring-shaped device using epitaxial Bi/Ni and the basic device property.** (**A**) Schematic image of Bi/Ni bilayer film epitaxially-grown on an MgO substrate. (**B**) Schematic image of the measurement setup. (**C**) Temperature dependence of resistance measured with a ring-shaped device (device #1). Inset: Scanning electron microscope image of device #1. (**D**) A picture of our sample holder. Bi/Ni (left) and Nb (right) devices were mounted on the same sample holder within 1 cm. The electrodes were attached to the sample holder with Al-Si (1%) bonding wires.

**Fig. 2. Resistance oscillations measured with epitaxial Bi/Ni and Nb ring-shaped devices.** (**A** and **B**) Resistance oscillations of epitaxial Bi/Ni (device #1) (**A**) and Nb (**B**) ring-shaped devices. The horizontal axis is the value obtained with a Hall sensor. Only for the epitaxial Bi/Ni device, there are two magnetic field regions ($\mu_0|H^*| \approx 7$ -9 mT, indicated by arrows) where the resistance oscillations are suppressed. The real zero field for the sample space is determined so that both resistance oscillations take minima and is identified as the vertical dashed line ($\approx -0.2$ mT). (**C**) Resistance oscillations for the epitaxial Bi/Ni ring-shaped device (device #1) at 2.47 K as a function of the magnetic field normalized by $\mu_0 H_0$. The black and red vertical lines indicate the integer and half-integer numbers of $H/H_0$, respectively. QFs-type oscillations are obtained in the former region, while HQFs-type oscillations are realized in the latter region. The crossover takes place at $\mu_0|H^*|$. (**D**)

Resistance oscillations for the epitaxial Bi/Ni ring-shaped device (device #1) at 3.6 K as a function of the magnetic field normalized by $\mu_0 H_0$. Only QFs-type oscillations are observed. We note that $I_\mathrm{m} = 17$ μA at 3.6 K is much lower than that at 2.47 K. That is the reason why the background resistance at 3.6 K is lower than that at 2.47 K. (**E**) Resistance oscillations for the epitaxial Bi/Ni ring-shaped device (device #1) at 2.47 K obtained with different $I_\mathrm{m}$. (**F**) Resistance oscillations for the epitaxial Bi/Ni ring-shaped device (device #2) at 2.46 K as a function of the magnetic field normalized by $\mu_0 H_0$.

**Fig. 3. Resistance oscillations measured with polycrystalline Bi/Ni and Nb/Ni ring-shaped devices.** (**A** and **B**) Resistance oscillations for polycrystalline Bi/Ni ring-shaped device (device #6) at $T = 2.45$ K (**A**) and Nb/Ni ring-shaped device (device #10) at $T = 2.41$ K (**B**), as a function of the magnetic field normalized by $\mu_0 H_0$. The black vertical lines indicate the integer numbers of $H/H_0$. Only QFs-type oscillations are obtained for both devices.

**Fig. 4. Calculated critical current density as a function of magnetic flux at different temperatures.** (**A**) Critical current as a function of magnetic flux at $T = 0.6 T_\mathrm{c}$. The difference in color represents the difference in the spin (**d**-vector) texture (see Sec. 9 in Supplementary Materials for more details). Among several critical currents for various types of spin textures, the highest one gives the physical critical current density at each field, as indicated by a black dotted curve. The inset shows the spin textures realized in the low-field and high-field (half-quantum-shifted) regions. There is a crossover from QF to HQF at $\Phi/\Phi_0 \approx 10$. (**B**) Critical current as a function of magnetic flux at $T = 0.9T_c$. No crossover is obtained.

Figure 1

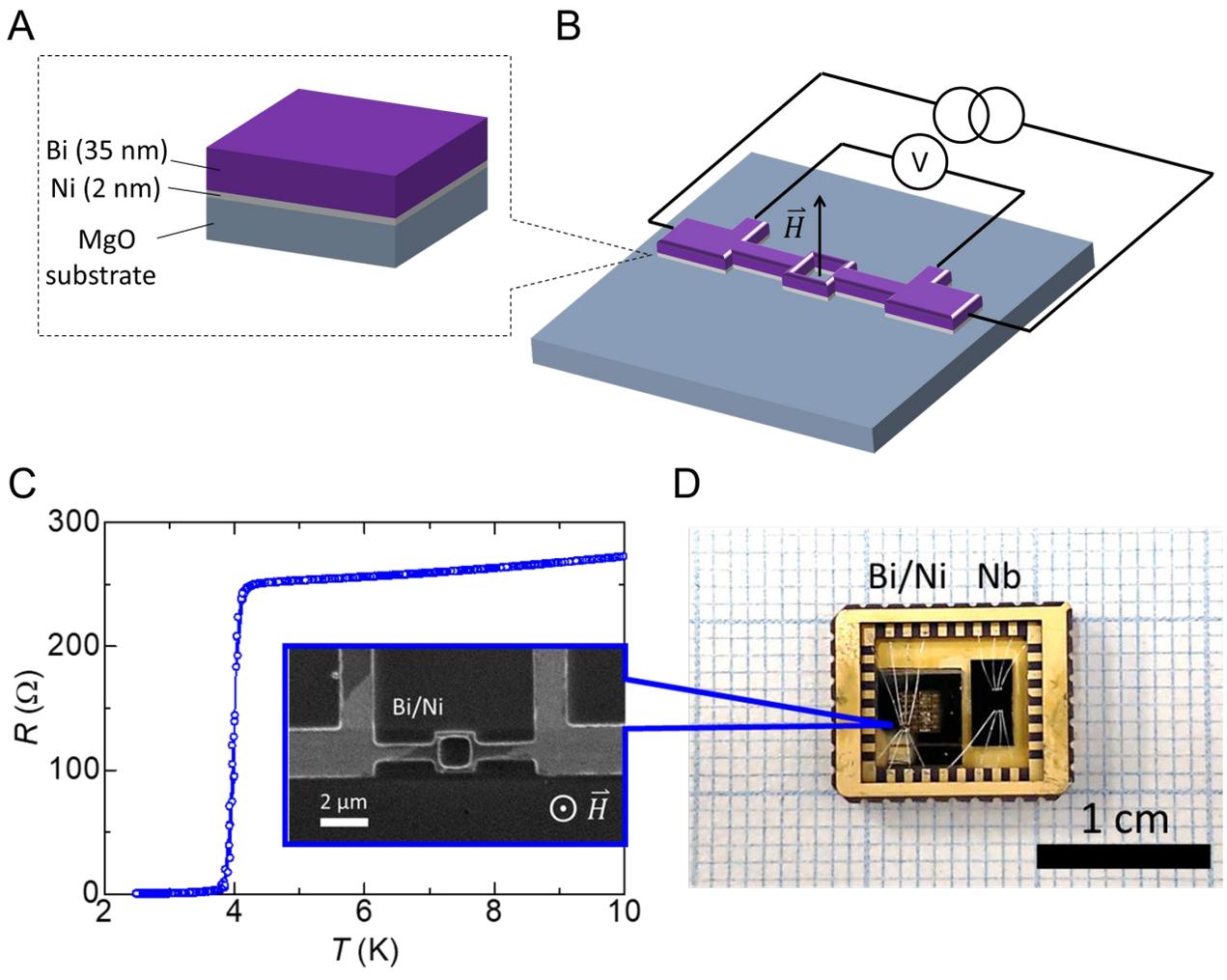

Figure2

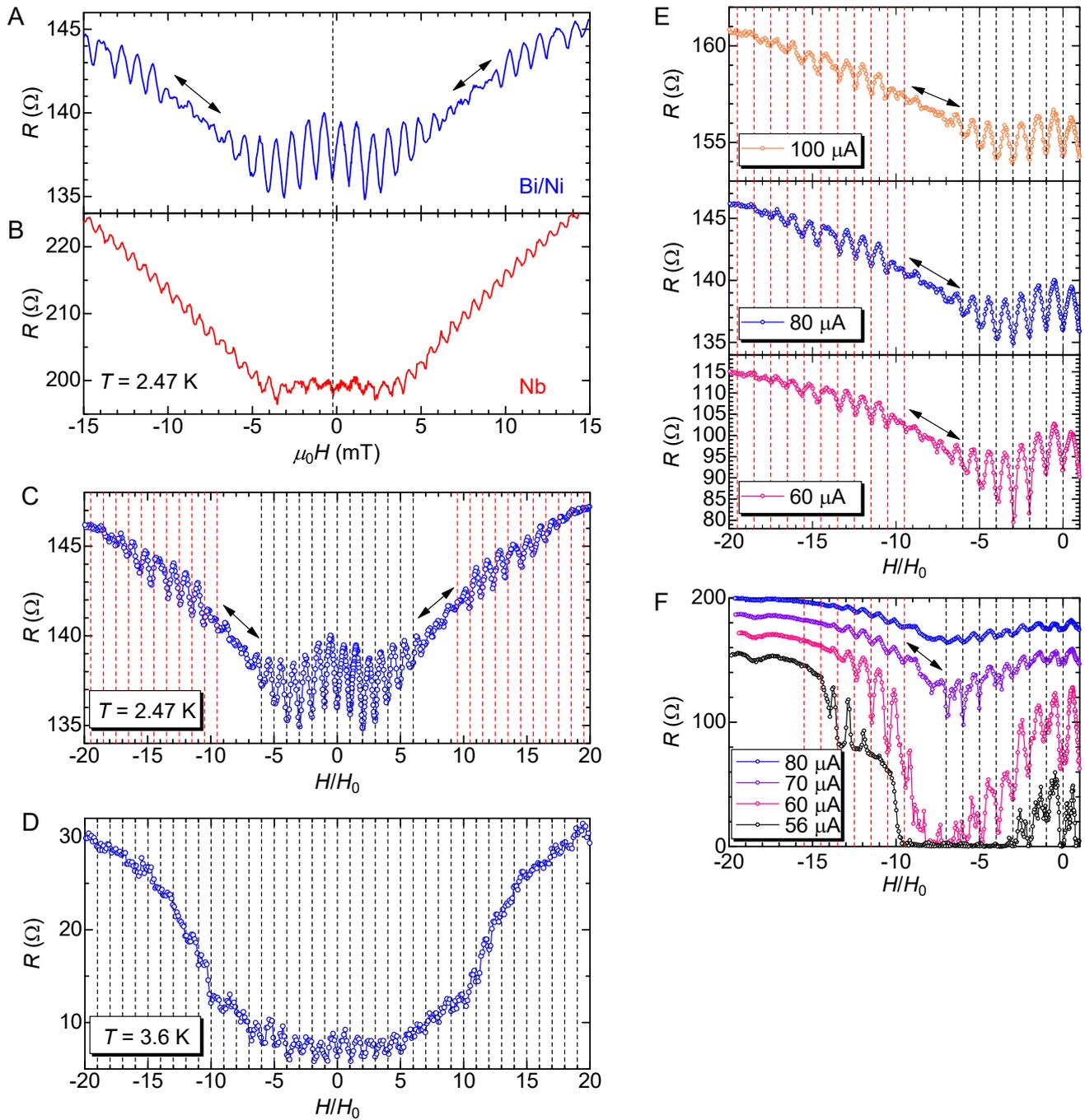

Figure 3

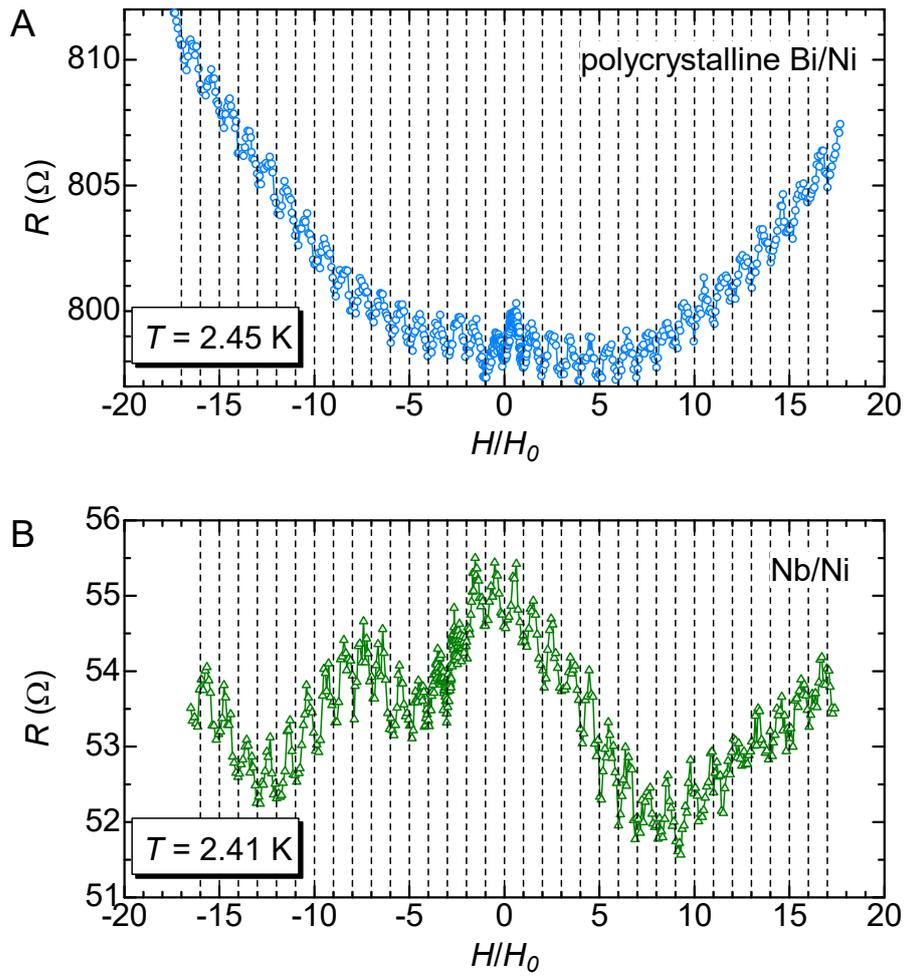

Figure4

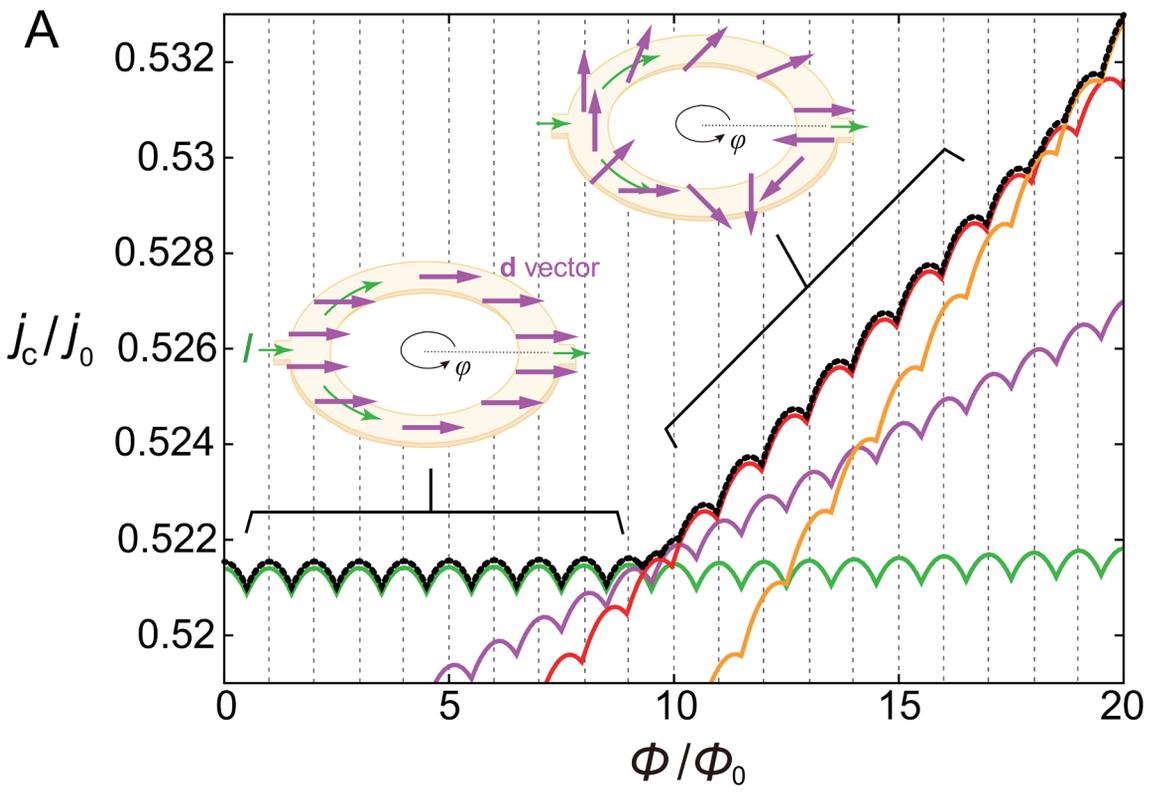

A

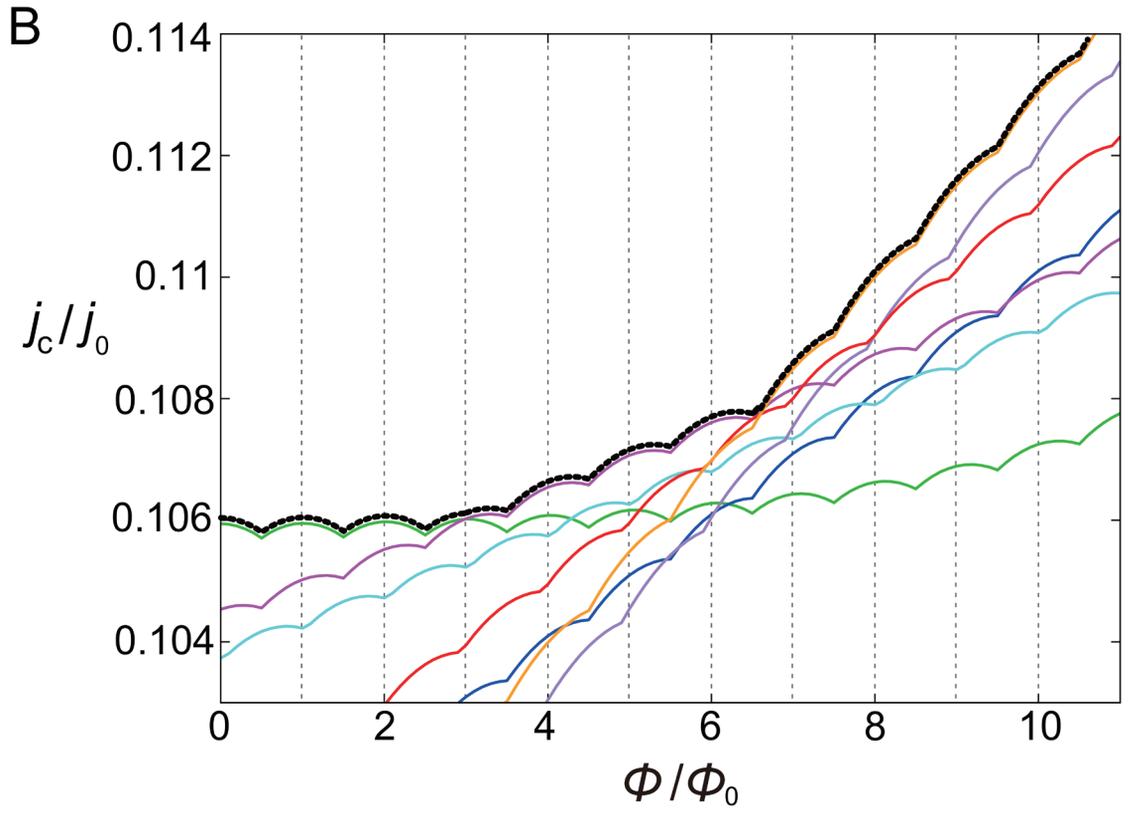

B